\documentclass[10pt,journal]{IEEEtran}
\makeatletter
\def\ps@headings{%
\def\@oddhead{\mbox{}\scriptsize\rightmark \hfil \thepage}%
\def\@evenhead{\scriptsize\thepage \hfil \leftmark\mbox{}}%
\def\@oddfoot{}%
\def\@evenfoot{}}
\makeatother
\usepackage{amsmath,amsthm}
\usepackage{enumitem}
\usepackage{amssymb}
\usepackage{fancyvrb}
\usepackage{graphicx,graphics}
\usepackage{algorithm}
\usepackage{algorithmic}
\usepackage{url}
\usepackage{balance}
\usepackage{longtable}
\usepackage{subcaption}
\usepackage{epstopdf}
\usepackage{tabularx}
\usepackage{boxedminipage}
\usepackage{color}
\newtheorem{theorem}{Theorem}
\newtheorem{lemma}[theorem]{Lemma}

\newtheorem{remark}[theorem]{Remark}

\allowdisplaybreaks

\begin{document}

\title{Distributed Rate and Power Control in Vehicular Networks}

\author{Jubin Jose, Chong Li, Xinzhou Wu, Lei Ying and Kai Zhu%
\thanks{J.Jose, C. Li and X. Wu are with Qualcomm Research; L,Ying and K. Zhu are with School of Electrical, Computer and Energy Engineering,
Arizona State University. The authors are listed in alphabetical order}}


\maketitle
\begin{abstract}
The focus of this paper is on the rate and power control algorithms in Dedicated Short Range Communication (DSRC) for vehicular networks. We first propose a utility maximization framework by leveraging the well-developed network congestion control, and formulate two subproblems, one on rate control with fixed transmit powers and the other on power control with fixed rates. Distributed rate control and power control algorithms are developed to solve these two subproblems, respectively, and are proved to be asymptotically optimal. Joint rate and power control can be done by using the two algorithms in an alternating fashion. The performance enhancement of our algorithms compared with a recent rate control algorithm, called EMBARC \cite{BanLuKen_13}, is evaluated by using the network simulator ns2.
\end{abstract}


\section{Introduction}
Dedicated Short Range Communication (DSRC) service \cite{kenney_jour11} is for vehicle-to-vehicle and vehicle-to-infrastructure communication in the $5.9$ GHz band. Among the $75$ MHz allocated to DSRC, the channel $172$ ($5.855$ GHz -- $5.865$ GHz) is assigned for critical safety operations, which allows vehicles to periodically exchange Basic Safety Messages ({\bf BSM}) to maximize the mutual awareness to prevent collisions. Such messages typically include the GPS position, velocity of the vehicle. By receiving these {\bf BSM} messages from surrounding vehicles, all participating vehicles in the DSRC safety system can assess the threat of potential collisions and provide warnings to the driver, if necessary. United States Department of Transportation (USDOT) reported that the DSRC safety system based on this simple mechanism can address $80\%$ of the traffic accidents on the road today and thus has potentially huge societal benefit. On the other hand, such benefit is possible only when timely and reliable information exchange among vehicles using DSRC can be guaranteed in \textsl{all} deployment scenarios.



DSRC is based on IEEE 802.11p standards \cite{80211p} at PHY and MAC layer. It has been well known that DSRC vehicular networks exhibit degraded performance in congested scenarios. 
In particular, excessive packet loss can be observed at high node density even between vehicles which are in close proximity to each other \cite{nguyen2013performance}, which can severely undermine the safety benefit targeted by deploying such networks. Performance improvement in such scenarios has been one of the key challenging issues for the success of DSRC. The industry and academics have contributed various solutions to this issue in a collaborative way over the last decade, e.g., \cite{etsi102687,BanKenRoh_13,krishna_control}.

The key system parameters one may control at each vehicle are the transmit power and transmit rate, i.e. the periodicity of the BSM messages, to alleviate the system congestion, i.e. lower transmit rate and power reduces the footprint and also the number of messages a DSRC device may generate and thus reduce the congestion level in the critical safety channel. On the other hand, both rate control and power control are critical for system performance as the transmit power of a vehicle determines the number of surrounding vehicles which can discover the vehicle and higher transmit rate of the BSM message can improve the accuracy the of collision estimation between two vehicles by having more message exchanges. Thus, a key problem to be addressed in the DSRC system is: \emph{How to choose the transmit rate and power for each vehicle in a distributed manner such that the overall system performance is maximized without creating excessive network congestion, i.e. observing very high channel load in some locations of the network?}

Both rate control and power control have been studied in the literature (e.g. \cite{NasFalKri_13,TieJiaHar_13}). 
However, most of these works propose \emph{heuristic} methods to adjust the rate and (or) power in simplistic scenarios, e.g. single bottleneck scenarios (i.e., there is only one congested channel in the network). Further, some of the methods \cite{BanKenRoh_13,BanLuKen_13} rely on the existence of global parameters for algorithm convergence, which leads to system resource \emph{under-utilization} in some scenarios.

The focus of this paper\footnote{Partial version of this paper has appeared in \cite{Jubin2015}.} is to propose a network resource allocation framework for rate and power control in vehicular network, by leveraging existing network congestion control framework \cite{SriYin_14} established in the context of wireline and wireless networks, and then develop optimal distributed rate and power control algorithms to achieve such a goal. The main contributions of this paper are summarized below:
\begin{itemize}
\item We propose a utility maximization framework for rate and power control in DSRC. In general, the utility maximization is a non-convex optimization problem with integer constraints. We separate the problem to two subproblems: rate control problem and power control problem, where the rate control problem is to find the optimal broadcast rates when the transmit power (i.e., the transmit ranges) of vehicles are fixed and the power control problem is to find the optimal transmit power (or transmission range) when the broadcast rates are fixed.

\item We develop a distributed rate control algorithm which is similar to the dual algorithm for the Internet congestion control and prove that the  time-average total utility obtained under the proposed rate control algorithm can be arbitrarily close to the optimal total utility with fixed transmission power.

\item The power control problem is a non-convex optimization problem. We reformulate the problem as an integer programming problem. After relaxing the integer constraints, we develop a distributed power control algorithm based on the dual formulation. Interestingly, it can be shown that one of the optimal solutions to the Lagrangian dual, when fixing the dual variables, is always an integer solution. That implies that the distributed algorithm derived from the relaxed optimization problem produces a valid power control decision and the relaxation is performed without loss of optimality (more details can be found in Section \ref{sec:alg_power}). Based on that, we prove that the time-average total utility obtained under the proposed power control algorithm can be arbitrarily close to the optimal total utility with fixed broadcast rates.

\end{itemize}

The paper is organized as follows. In Section \ref{sec:formulation}, a utility maximization framework is provided for congestion control in DSRC. Then asymptotically optimal distributed rate control algorithm and power control algorithm are derived respectively in Section \ref{sec:alg_rate} and \ref{sec:alg_power}. In the end, simulation results are presented in Section \ref{sec: evaluation} to verify our algorithms. The proofs in the paper are provided in the Appendix.

\subsection{Discussion on Related Work}
The design of rate and power control algorithms in DSRC is one of most critical problems in ITS. Error Model Based Adaptive Rate Control (EMBARC) \cite{BanLuKen_13} is a recent rate control protocol which integrates several existing rate control algorithms including the Linear Integrated Message Rate Control (LIMERIC) \cite{BanKenRoh_13}, Periodically Updated Load Sensitive Adaptive Rate control (PULSAR) \cite{TieJiaChe_11}, and the InterVechicle Transmission Rate Control (IVTRC) \cite{HuaFalSen_11}.  LIMERIC allocates the wireless channel {\em equally} among all vehicles that share the same bottleneck link  while guaranteeing the channel load is below a given threshold. IVTRC generates messages and adapts transmission probabilities based on the Suspected Tracking Error (STE) calculated based on vehicle dynamics to avoid collisions. In EMBARC, the message rates are controlled by LIMERIC and are further modified to satisfy the STE requirement.

A parallel work \cite{Zhang_safety2014} introduced a network utility maximization (NUM) formulation on the rate control problem when specified to safety-awareness. A distributed algorithm was proposed to adjust the rate with the objective to maximize the utility function. Similarly, \cite{egea2014distributed} also provided a NUM formulation on the rate control problem and proposed a fair adaptive beaconing rate for intervehicular communications (FABRIC) algorithm, which essentially is a particular scaled gradient projection algorithm to solve the dual of the NUM problem.


Other related work includes the database approach proposed in \cite{TieJiaHar_13}, where the optimal broadcast rates and transmission power are calculated offline based on the network configurations. Also, \cite{Bengi15} proposed an environment and context-aware distributed congestion control (DCC) algorithm, which jointly control the rate and power to improve cooperative awareness by adapting to both specific propagation environments (such as urban intersections, open highways, suburban roads) as well as application requirements (e.g., different target cooperative awareness range). However, the stability and convergence of the algorithm are not proved mathematically. Besides the rate control algorithm IVTRC, the authors also proposed range control algorithms in \cite{FalHuaSen_10,HuaFalSen_10,NasFalKri_13} where the objective is to adapt the transmission ranges to achieve a specific threshold. The motivation of limiting channel loads below the threshold is to control channel congestion to maximize effective channel throughput. However, fair resource allocation among vehicles to increase the safety awareness of all vehicles are not considered, and the stability of the algorithms is subject to certain conditions \cite{NasFalKri_13}.

\section{Problem Formulation}
\label{sec:formulation}
In this section, we formally define the utility maximization framework for the DSRC congestion control problem. We first introduce the set of notations used throughout this paper.
\begin{itemize}
\item $\mu_i:$ the message broadcast rate of vehicle $i;$
\item $p_i:$ the transmit power of vehicle $i;$
\item $\alpha_{ij}:$ the minimum transmit power required for node $i$'s message to be \emph{decoded} by node $j;$
\item $\beta_{ij}:$ the minimum transmit power required for node $i$'s message to be \emph{sensed} by node $j,$ i.e. the received energy is above the carrier sensing energy detection threshold;
\item $\mathbb I:$ indicator function.
\end{itemize}
Note $\alpha_{ij}$ is not necessarily the same as $\beta_{ij},$ as in IEEE802.11 standards, packet header decoding happens at a much lower energy level than energy based detection in carrier sensing. From the definition of $\alpha_{ij}$ and $\beta_{ij},$ it is easy to see that
\begin{itemize}
\item Vehicle $j$ can receive the message from vehicle $i$ if $p_{i}\geq \alpha_{ij};$
\item Vehicle $j$ can detect (but not necessarily decode) the message from vehicle $i$ if $p_i\geq \beta_{ij}.$
\end{itemize}
We assume $\alpha_{ij}$ and $\beta_{ij}$ are constants within the time frame for the distributed rate and power control algorithm, which is reasonable as the nominal BSM update rate is $10$Hz, i.e. $10$ transmissions in every second.

In DSRC, a vehicle can control rate $\mu_i$ and power $p_i.$ We consider the following utility maximization problem for rate and power control:
\begin{eqnarray}
{\bf General-OPT}&\max_{\mu, {\bf p}}\sum_{i} \sum_j {\mathbb I}_{p_i\geq \alpha_{ij}} U_{ij}\left({\mu_i}\right) \label{opt:obj0}\\
\hbox{subject to:}& \sum_i \mu_i {\mathbb I}_{p_i\geq \beta_{ij}}\leq \gamma\quad \forall j \label{opt:const0}\\
&\mu_i\geq 0, \ p_i\geq 0\quad \forall i.
\end{eqnarray}

Now we explain the particular choice of the objective function and constraints above. In the objective function (\ref{opt:obj0}), $$ \sum_j {\mathbb I}_{p_i\geq \alpha_{ij}} U_{ij}\left({\mu_i}\right)$$ is the total utility associated with vehicle $i,$ which depends on the number of vehicles who can receive the transmissions of vehicle $i,$ i.e., the size of the set \begin{equation}\left\{j: {\mathbb I}_{p_i\geq \alpha_{ij}}=1\right\}\label{eq:nsize}\end{equation} and the utility function $U_{ij}(\mu_i)$ associated with each ordered pair $(i,j),$ which is a concave function and can be interpreted as the level of $j'$s awareness of $i$ when $j$ receives messages from $i$ with rate $\mu_i.$ Obviously, higher transmission rate $\mu_i$ should lead to higher value of $U_{ij}$ in DSRC. The neighborhood size (\ref{eq:nsize}) is controlled by the transmit power $p_i$ and the value of utility $U_{ij}(\mu_i)$ is determined by rate $\mu_i.$

\begin{remark}
A widely used utility function \cite{FairEndtoEnd2000} is called the $\alpha$-fair utility function which includes proportional fairness, minimum potential-delay fairness as special cases and is given by
\begin{equation}
U_{i}(\mu_i) = w_i\frac{\mu_i^{1-\alpha_i}}{1-\alpha_i}, \quad \alpha_i>0,
\end{equation}
where $w_i$ represents the weight of node $i$, determined by its local information such as relative velocity, instantaneous location in the application of vehicular network. Notice that this utility function, given in a generic form, is independent of communication links (from $j$). In other words, each vehicle only knows its own utility function. As will be seen later, a choice of such a form of utility function further simplifies the proposed distributed algorithms because there is no need of obtaining neighbors' utility functions.

For $\alpha_i = 2$, the utility function turns to be $U_{i}(\mu_i) = -\frac{w_i}{\mu_i}$ which implies weighted minimum potential delay fairness in network's resource allocation. For $\alpha_i = 1$, the utility function behaves as $U_{i}(\mu_i) = w_i\log(\mu_i)$ which leads to weighted proportional fairness. We refer interested readers to \cite{Sri_theMathematics12} for details.
\end{remark}

The constraint (\ref{opt:const0}) states that the \emph{channel load} at any vehicle $j$ should be below a target channel load $\gamma.$ In CSMA based systems, high $\gamma$ value indicates channel congestion, which implies high packet collision rate and low \emph{effective} channel throughout \cite{FalHuaSen_10,TieJiaHar_13}.
In \cite{FalHuaSen_10}, the authors have observed that the curve of information dissemination rate versus channel load remains almost the same under different configurations. In \cite{TieJiaHar_13}, the authors also found that the effective channel throughput is maximized when the channel load is around $0.91$ under various settings. Thus, it is natural to impose such a constraint (\ref{opt:const0}) to limit the congestion level in the system.

\subsection{Problem decomposition}

General-OPT is difficult to solve because 
the objective function (\ref{opt:obj0}) is \emph{not} jointly convex in $(\mu,{\bf p}).$ We therefore separate the problem into rate control problem and power control problem as defined below.
\begin{itemize}
\item Assume the transmit power is fixed at each vehicle. Then we can define $${\cal R}_i=\{j: p_i\geq \alpha_{ij}\},$$ i.e., the set of vehicles who can receive the messages from vehicle $i,$  and
$${\cal I}_i=\{j: p_i\geq \beta_{ij}\},$$ i.e., the set of vehicles whose channel load can be affected by vehicle $i'$s transmissions. When transmit power $p_i$ is fixed, both ${\cal R}_i$ and ${\cal I}_i$ are fixed. In this case, general-OPT becomes the following Rate-OPT
\begin{eqnarray}
{\bf Rate-OPT:} \quad \rho = &\max_{\mu}\sum_i \sum_{j\in{\cal R}_i} U_{ij}({\mu_i}) \label{rateopt:obj}\\
\hbox{subject to:}& \sum_{i: j\in{\cal I}_i} \mu_i \leq \gamma \quad \forall \ j.\label{rateopt:cons}
\end{eqnarray}

\item Assuming the broadcast rates are fixed, i.e., $\mu_i$'s are fixed, OPT becomes the following Power-OPT:
\begin{eqnarray}
{\bf Power-OPT:} \quad \rho = &\max_{{\bf p}}\sum_{i,j} {\mathbb I}_{p_i\geq \alpha_{ij}} U_{ij}(\mu_i) \label{rangeopt:obj}\\
\hbox{subject to:}& \sum_{i} \mu_i {\mathbb I}_{p_i\geq \beta_{ij}} \leq \gamma.
\end{eqnarray}
\end{itemize}

\subsection{Iterative joint rate and power control}

In light of the above decompositions, a (suboptimal) solution of General-OPT can be obtained by iterating Rate-OPT and Power-OPT. The initial set of rate or power parameters for the iterative algorithm can be appropriately chosen according to certain practical constraints. The \textit{stopping criterion} at step $k$ is typically set to be $\rho(k+1) - \rho(k)\leq \epsilon$ for $\epsilon>0$. It is worth noting that in each step of iterations the utility value $\rho(k)$ is non-decreasing and $\rho(k)$ is bounded above for all $\forall k$, given a well-defined utility function. Therefore, the convergence of the iterative algorithm is guaranteed.

In the following sections, we will develop distributed algorithms to solve Rate-OPT and Power-OPT separately. The optimal rate control algorithm directly follows from the well-developed network congestion control while the optimal power control algorithm is innovative and rather technical.


\section{Rate Control Algorithm}
\label{sec:alg_rate}
In what follows, we study the rate control problem and develop a distributed rate control algorithm that solves (\ref{rateopt:obj}).

Note that Rate-OPT is similar to the network utility maximization (NUM) problem for the Internet congestion control (see \cite{SriYin_14} for a comprehensive introduction of the NUM problem for the Internet congestion control). Each vehicle $i$ may represent both a flow and a link on the Internet, and $\mu_i$ is the data rate of flow $i.$ Regarding $\sum_{j\in{\cal R}_i} U_{ij}({\mu_i})$ as the utility function of vehicle $i,$  the objective is to maximize the sum of user utilities. We may further say that flow $i$ uses link $j$ when $j\in {\cal I}_i.$  Then constraint (\ref{rateopt:cons}) is equivalent to the link capacity constraint that requires the total data rate on link $j$ to be no more than the link capacity $\gamma.$ To this end, Rate-OPT can be viewed as a standard NUM problem for the Internet congestion control. The distributed rate control algorithm below is based on the dual congestion control algorithm for the Internet congestion control \cite{SriYin_14}, which consists of rate control and congestion price update. The congestion price update monitors the channel load of vehicle $j.$ The congestion price $\lambda_j$ increases when the channel load at vehicle $j$ exceeds the threshold $\gamma$ and decreases otherwise. The rate control algorithm adapts the broadcast rate $\mu_i$ based on the aggregated congestion price from all vehicles {\em who can sense the transmissions from vehicle $i$,} i.e., the vehicles whose channel loads are affected by vehicle $i.$

\vspace{0.1in}
\hrule
\vspace{0.1in}

\noindent{\bf Rate Control Algorithm}

\begin{enumerate}[leftmargin=*]
\item Rate control algorithm at vehilce $i:$ At time slot $t,$ vehicle $i$ broadcasts with rate $\mu_i[t]$ such that
\begin{equation}
\begin{split}
&\mu_i[t]\\
=&\min\left\{\arg\max_{\mu} \sum_{j\in{\cal R}_i} U_{ij}({\mu})-\epsilon \mu\sum_{j\in{\cal R}_i} \lambda_j[t-1], \mu_{\max}\right\}
\end{split}
\label{alg:rate}
\end{equation}
where $\epsilon \in (0,1]$ is a tuning parameter.

\item Congestion price update at vehicle $j:$ At time slot $t,$ vehicle $j$ updates its congestion price $\lambda_j$ to be
\begin{equation}\lambda_j[t]=\left(\lambda_j[t-1]+\sum_{i:j\in{\cal I}_i}\mu_i[t-1]-\gamma\right)^+.\label{alg:price}\end{equation}
\end{enumerate}

\vspace{0.1in}
\hrule
\vspace{0.1in}

This rate control algorithm is developed based on the dual decomposition approach \cite{SriYin_14}. Specifically, the Lagrangian of optimization (\ref{rateopt:obj}) is
\begin{equation*}
\begin{split}
&\mathcal{L}(\mu_i,{\bf \lambda}) \\
=&\sum_{i}\sum_{j\in{\cal R}_i} U_{ij}({\mu}_i) -\epsilon \sum_j \lambda_j \left(\sum_{i: j\in{\cal I}_i} \mu_i -\gamma\right)\\
=&\sum_{i} \left(\sum_{j\in{\cal R}_i} U_{ij}({\mu}_i) -\epsilon \mu_i \sum_{j\in{\cal I}_i} \lambda_j\right) -\gamma \sum_j \lambda_j,
\end{split}
\end{equation*}
where $\epsilon$ is a tuning parameter. Then the dual problem is
$$ \min_{{\bf \lambda}} g({\bf \lambda}) =  \min_{{\bf \lambda}} \max_{\mu_i} \mathcal{L}(\mu_i,{\bf \lambda})$$
When $\lambda$ is fixed, the $\mu_i$ should maximize
$$\sum_{j\in{\cal R}_i} U_{ij}({\mu}_i) -\epsilon \mu_i \sum_{j\in{\cal I}_i} \lambda_j,$$
which motivates the rate control algorithm (\ref{alg:rate}). The congestion price update (\ref{alg:price}) is designed by taking derivative of $g({\bf \lambda})$ over ${\bf \lambda}$ and then the optimal ${\bf \lambda}$, as a mean to optimize the dual problem, can be achieved by using a gradient search in (\ref{alg:price}).

The next theorem shows the rate control algorithm is asymptotically optimal.

\begin{theorem}
Denote by $\mu_i^*$ the optimal solution to problem (\ref{rateopt:obj}) and assume $\mu_{\max}>\mu_i^*$ for all $i.$ Then there exists a constant $B>0,$ independent of $\epsilon,$ such that under the proposed rate control algorithm $$\liminf_{T\rightarrow\infty}\frac{1}{T}\sum_{t=0}^{T-1} \sum_i\sum_{j\in{\cal I}_i} U_{ij}({\mu}_i[t])\geq\sum_i \sum_{j\in{\cal I}_i} U_{ij}(\mu_i^*) -B\epsilon.$$ \hfill{$\square$}
\end{theorem}
 The proof of the theorem is similar to the proof of Theorem 6.1.1 in \cite{SriYin_14}, and is omitted in this paper. Remark that if the objective function utility function $\sum_{j\in{\cal R}_i} U_{ij}({\mu}_i)$ is \textit{strictly} concave, the optimal solution of Rate-OPT is \textit{unique} since the search space is convex. As a consequence, the above algorithm converges to the unique optimal solution.

\section{Power Control Algorithm}
\label{sec:alg_power}
In this section, we develop a distributed power control algorithm that solves (\ref{rangeopt:obj}). The power control problem is developed by formulating the Power-OPT as an integer programming problem. After relaxing the integer constraint, we develop a distributed power control algorithm using the dual approach. Interestingly, it turns out the solution obtained from the Lagrangian dual is always an integer solution. In other words, the power control algorithm based on the linear approximation always gives a valid power control solution and is proved to be asymptotically optimal for Power-OPT.

We first introduce new variables $x$ and $y$ such that $$x_{ij}={\mathbb I}_{p_i\geq \alpha_{ij}}\quad\hbox{and}\quad y_{ij}={\mathbb I}_{p_i\geq \beta_{ij}}.$$ The Power-OPT problem is equivalent to the following integer programming problem:
\begin{eqnarray}
&\max_{x,y}\sum_{i,j} x_{ij} U_{ij}(\mu_i)& \label{eq:obj}\\
\hbox{subject to:}& \sum_{i} y_{ij}\mu_i \leq \gamma & \forall \ j\label{cons:load}\\
&x_{ij}\geq x_{ik} & \forall \ \alpha_{ij}\leq \alpha_{ik}\label{cons:range}\\
&y_{ik}\geq x_{ij} & \forall \ \beta_{ik}\leq \alpha_{ij}\label{cons:sen}\\
& x_{ij}\in \{0, 1\} &\forall i, j\label{cons:int}\\
& y_{ij}\in \{0, 1\} &\forall i, j.\label{cons:int2}
\end{eqnarray}
Recall $\alpha_{ij}$ is the minimum transmit power for vehicle $j$ to receive messages from vehicle $i.$ So constraint (\ref{cons:range}) states that if vehicle $j$ requires a smaller minimum transmit power of vehicle $i$ than vehicle $k,$ then vehicle $j$ can receive from vehicle $i$ if vehicle $k$ can do so.  Constraint (\ref{cons:sen}) is similarly defined.

Next, we relax the integer constraints (\ref{cons:int}) and (\ref{cons:int2}) to obtain the following linear programming problem.
\begin{eqnarray}
&\max_{x, y}\sum_{i,j} x_{ij} U_{ij}(\mu_i) \label{eq:obj1}&\\
\hbox{subject to:}& \sum_{i} y_{ij}\mu_i \leq \gamma  & \forall \ j \label{cons:load2}\\
&x_{ij}\geq x_{ik} & \forall \ \alpha_{ij}\leq \alpha_{ik}\label{cons:range2} \\
&y_{ik}\geq x_{ij}& \forall \ \beta_{ik}\leq \alpha_{ij}\label{cons:sen2}\\
&0\leq x_{ij}\leq 1& \forall i, j\\
&0\leq y_{ij}\leq 1& \forall i, j.
\end{eqnarray}

Now by including constraint (\ref{cons:load2}) in the Lagrangian, we obtain
\begin{eqnarray*}
&\max_{x,y}\sum_{i,j} x_{ij} U_{ij}(\mu_i) -\epsilon \sum_j \lambda_j \left(\sum_{i} y_{ij}\mu_i -\gamma\right)\\
\hbox{s.t.:}&x_{ij}\geq x_{ik} \quad \forall \ \alpha_{ij}\leq \alpha_{ik} &\\
&y_{ik}\geq x_{ij} \quad \forall \ \beta_{ik}\leq \alpha_{ij}&\\
&0\leq x_{ij}\leq 1\quad \forall i, j &\\
&0\leq y_{ij}\leq  1\quad \forall i, j.&
\end{eqnarray*}
where $\epsilon$ is a tuning parameter. Note that constraints (\ref{cons:range2}) and (\ref{cons:sen2}) impose conditions on $x$ and $y$ related to the same transmitter $i.$ Therefore, given $\lambda,$  the Lagrangian dual problem above can be decomposed into the sub-problems for each given $i:$
\begin{eqnarray}
&\max_{x, y}\sum_{j} x_{ij} U_{ij}(\mu_i) -\epsilon \lambda_j\mu_i y_{ij}\label{subopt:obj}\\
\hbox{subject to:}&x_{ij}\geq x_{ik} \quad \forall \ \alpha_{ij}\leq \alpha_{ik} &\nonumber\\
&y_{ik}\geq x_{ij} \quad \forall \ \beta_{ik}\leq \alpha_{ij}&\nonumber\\
&0\leq x_{ij}\leq 1\quad \forall j &\nonumber\\
&0\leq y_{ij}\leq  1\quad \forall j.&\nonumber
\end{eqnarray}

Next we will show that one of the optimizers to the problem (\ref{subopt:obj}) is an integer solution. For a fixed vehicle $i,$ we sort the vehicles in a descendent order according to $\alpha_{ij}$ and divide them into groups, called G-groups and denoted by ${\cal G}_g,$ such that $\alpha_{ij}=\alpha_{ik}$ if $j, k\in {\cal G}_g,$ and $\alpha_{ij}<\alpha_{ik}$ if $j\in {\cal G}_g$ and $k\in {\cal G}_{g+1}.$ Associated with each group ${\cal G}_g,$ we define $\tilde{\alpha}_g$ to be common $\alpha$ in the group. We further define H-groups
$${\cal H}_g=\{m: \tilde{\alpha}_{g-1}< \beta_{im}\leq \tilde{\alpha}_g \}.$$ This is the set of vehicles that can sense the transmission of vehicle $i$ when the transmit power is $\tilde{\alpha}_g$ and cannot if the power is $\tilde{\alpha}_{g-1}.$ Furthermore, let $g(j)$ denote the G-group vehicle $j$ is in and $h(j)$ the H-group vehicle $j$ is in. The following lemma proves that one of the optimal solution to (\ref{subopt:obj}) is an integer solution. The proof is presented in the Appendix.

\begin{lemma}
Given $\lambda,$ one of optimizers to optimization problem (\ref{subopt:obj}) for given vehicle $i$  is the following integer solution
$$x_{ij}=\left\{
           \begin{array}{ll}
             1, & \hbox{if } g(j)\leq g^\prime_i  \\
             0, & \hbox{otherwise.}
           \end{array}
         \right. \quad\hbox{and}\quad y_{ij}=\left\{
           \begin{array}{ll}
             1, & \hbox{if } h(j)\leq g^\prime_i  \\
             0, & \hbox{otherwise.}
           \end{array}
         \right.,
$$ where
\begin{align*}
g^\prime_i=\max&\left\{g: \sum_{j\in\cup_{q:k\leq q\leq g}{\cal G}_{q}}U_{ij}(\mu_i)\right.\\
&\left.-\epsilon\sum_{j\in\cup_{q:k\leq q\leq g}{\cal H}_{q}}\lambda_j\mu_i>0\quad \forall 0\leq k\leq g\right\}.\vspace{-2em}
\end{align*}
\hfill{$\square$}
\label{lem:integer}
\end{lemma}

\begin{algorithm}
\caption{Sample algorithm for Lemma \ref{lem:integer}}
\begin{algorithmic}[1]
\REQUIRE{$g_{max}$, $\lambda_j$.}
\ENSURE{$g^\prime_i$}
\STATE{Define $f_p = \sum_{j\in{\cal G}_{p}}U_{ij}(\mu_i)-\epsilon\sum_{j\in{\cal H}_{p}}\lambda_j\mu_i, p = 1,2,\cdots, g_{max}$. }
\FORALL{$g \in\lbrace g_{max},g_{max}-1, \cdots, 1\rbrace$}
\STATE{$k \leftarrow g$ and $flag \leftarrow 1$.}
\WHILE{$flag = 1$ and $k\geq 0$}
\STATE{$F_g \leftarrow \sum_{p = k}^{g} f_p$;}
\IF{$F_g > 0$}
\STATE{$k \leftarrow k-1 $}
\ELSE
\STATE{$flag = 0$}
\ENDIF
\ENDWHILE
\IF{$flag = 1$}
\STATE{$g^\prime_i \leftarrow g $}; break;
\ENDIF
\ENDFOR
\end{algorithmic}
\label{alg_bisection_distributed}
\end{algorithm}

The optimization in Lemma \ref{lem:integer} can be solved by low complexity algorithms. A sample algorithm is given in Algorithm \ref{alg_bisection_distributed}. Note that the optimization problem can be further simplified for specific utility functions, e.g., $U_{i,j}(\mu_i) = w_i \log(\mu_i)$.

Based on the discussion and lemma above, we develop the following power control algorithm, which consists of congestion price update and power update. The congestion price update monitors the channel load and the power update adapts the transmission power $p_i$ based on the aggregated congestion price from all vehicles {\em who can sense the transmissions from vehicle $i.$}

\vspace{0.1in}
\hrule
\vspace{0.1in}

\noindent{\bf Power Control Algorithm}

\begin{enumerate}[leftmargin=*]
\item Power control at vehicle $i:$ Vehicle $i$ chooses the transmission power to be
\begin{equation}
p_i[t+1]=\tilde{\alpha}_{g^\prime_i},
\label{alg:power_01}
\end{equation} where $g^\prime_i$ is defined in Lemma \ref{lem:integer} with $\lambda=\lambda[t].$

\item Congestion price update at vehicle $j:$
\begin{equation}
\lambda_j[t+1]=\left(\lambda_j[t]+\sum_{i:j\in{\cal I}_i} \mu_i-\gamma\right)^+.\label{alg:price-r}
\end{equation}

\end{enumerate}

\vspace{0.1in}
\hrule

\begin{remark}
Notice that the second step of the power control, congestion price update, is identical to that in the rate control. In practice, the value of $\sum_{i:j\in{\cal I}_i} \mu_i$ can be approximated by measured/sensed channel load of individual vehicle\footnote{The channel load of DSRC is measured by carrier sensing technique which is widely implemented in CSMA network}. Furthermore, as shown in Lemma \ref{lem:integer}, the congestion prices of vehicles in the sensing range ${\cal I}_i$ are required in the power control (\ref{alg:power_01}) while only the prices of vehicles in the receiving range ${\cal R}_i$ are needed in the rate control. Unlike the price acquisition in the receiving range, which can be piggybacked in the broadcasted BSM, the price information of vehicles in the sensing range cannot be decoded. The approach of obtaining congestion prices in the sensing range is not discussed in this paper since it is rather implementation-specific and out of the scope of this paper.
\end{remark}
The next theorem shows the asymptotic optimality of the proposed distributed power control algorithm.

\begin{theorem}
Denote by $p^*$ the optimal solution to Power-OPT. There exists a constant $M>0,$ independent of $\epsilon,$ such that under the proposed power control algorithm,
\begin{align*}
&\liminf_{T\rightarrow \infty}\frac{1}{T}\sum_{t=0}^{T-1} \sum_{i,j} {\mathbb I}_{p_i[t]\geq \alpha_{ij}} U_{ij}(\mu_i) \geq \sum_{i,j} {\mathbb I}_{p^*_i\geq \alpha_{ij}} U_{ij}(\mu_i) -\epsilon M.
\end{align*}
\label{thm:range-opt} \hfill{$\square$}
\end{theorem}

\section{Performance Evaluation using ns2}
\label{sec: evaluation}

In this section, we evaluate the performance of the distributed rate and power control algorithm developed in this paper, and compare the performance with EMBARC \cite{BanLuKen_13}. We used the ns2 platform to simulate the asynchronous IEEE 802.11p media access control algorithm with the associated lower layer functions. The 802.11MacExt package in ns2 is adopted.

\begin{figure}
\begin{centering}
  \includegraphics[width=3.5in]{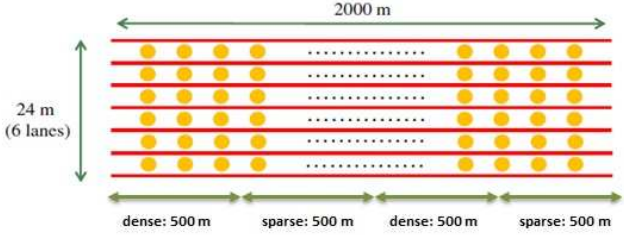}
  \caption{Deployment of $6$-lane highway vehicles}\label{figure:layout}
  \end{centering}
\end{figure}

To simulate congestion at high densities, we constructed a $6$-lane scenario where each lane is $4$ meters wide and $2000$ meters long. We use a wrap-around model of a network along the length of the road (see Figure \ref{figure:layout}). In each lane, $300$ vehicles are deployed in a {\em dense-sparse-dense-sparse} fashion as a grid. Specifically, the first $120$ vehicles are spaced with either $4$ or $5$ meters distance between any adjacent vehicles. Similarly, the next $30$ vehicles are spaced with either $16$ or $17$ meters distance.  The last $150$ vehicles are deployed by following the same rule. A comprehensive list of simulation parameters is summarized in Table \ref{tab:sim_para}.

\begin{table}[H]
\begin{center}
\begin{tabular}{|r|l|}
\hline
number of vehicles & 1800\\
\hline
packet size & 300 Byte \\
\hline
carrier frequency & 5.9 GHz\\
\hline
noise floor & -96 dBm \\
\hline
carrier sense threshold & -76 dBm\\
\hline
contention window & 15 \\
\hline
transmission rate & 6 Mbps\\
\hline
carrier sensing period & 0.25 s \\
\hline
\end{tabular}
\end{center}
\caption{Simulation Parameters} \label{tab:sim_para}
\end{table}

\begin{figure}
\begin{centering}
  \includegraphics[width=3in]{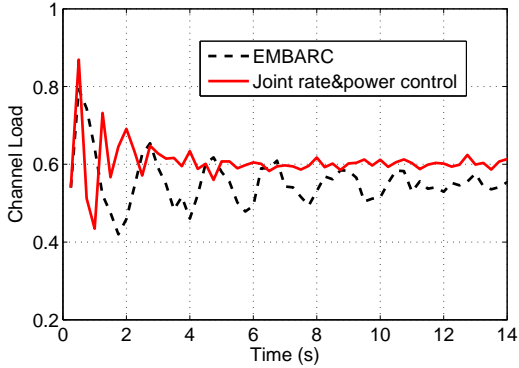}
  \caption{Convergence of channel load}\label{figure:channelload_compare}
  \end{centering}
\end{figure}

\begin{figure}[htb]
\begin{centering}
  \includegraphics[width=3in]{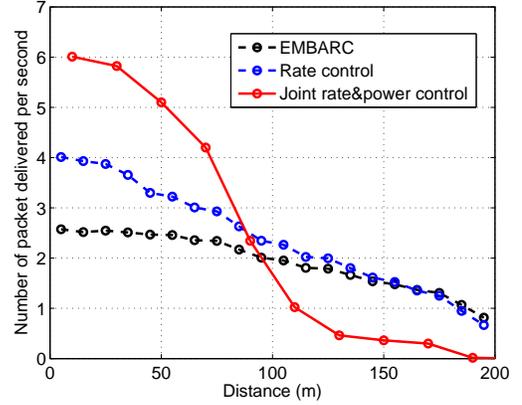}
  \caption{Number of successful received packets per second v.s. distance between transmitter and receiver}\label{figure:NumOf_Received_Packet}
  \end{centering}
\end{figure}

\begin{figure}
\begin{centering}
  \includegraphics[width=3in]{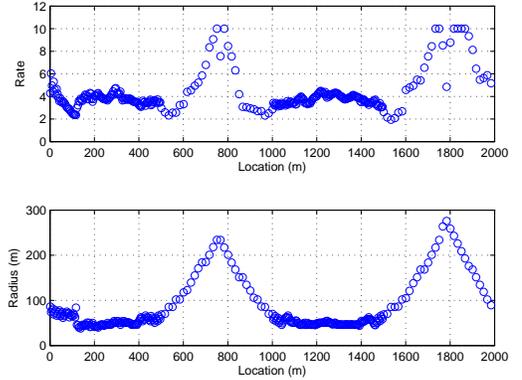}
  \caption{Broadcast rates and transmission ranges of the vehicles in the first lane under the joint rate and power control algorithm }\label{figure:Joint_performance}
  \end{centering}
\end{figure}

We now briefly review the EMBRAC algorithm, of which the transmission rate is a function of both channel load (using LIMERIC component) and vehicle dynamics (using the suspected tracking error component). In our ns2 simulations, we did not consider vehicle dynamics and assumed that the relative positions of the vehicles are static, which can be justified using a time-scale separation assumption under which the dynamics of the rate and power control algorithms are at a much faster time scale than the relative dynamics of the vehicles. Therefore, the suspected tracking error component of EMBARC was not simulated and EMBARC turns to be LIMERIC. According to \cite{BanLuKen_13}, LIMERIC is a distributed and linear rate-control algorithm and the rate of vehicle $i$ is evolving as follows,
\begin{equation}
r_i(t) = (1-\alpha) r_i(t-1) +\beta(r_g-r_c(t-1)),
\label{equ:LIMERIC}
\end{equation}
where $r_c$ is the total rate of all the $K$ vehicles and $r_g$ is the target total rate. $\alpha$ and $\beta$ are parameters that tunes stability, fairness and steady state convergence. In EMBARC, however, $r_c$ is defined to be the maximum channel load reported by all the 2-hop neighbors in order to achieve global fairness \cite{BanLuKen_13}.

For the implementation of our rate and power control algorithm, the sum rate from the interfering vehicles in the congestion price update equations (\ref{alg:price}) and (\ref{alg:price-r}) can be replaced by the measured channel load at vehicle $j.$ Therefore, each vehicle only needs to piggyback its congestion price in the safety message in broadcasting. Further, we chose the following utility function for evaluation
\begin{equation}
U_{ij}(\mu_i)= \frac{\max\{v_{ij},\alpha\}}{d_{ij}}\log {\mu_i}.\label{eq: utility.example}
\end{equation}
This specific choice of utility functions is motivated from the collision avoidance perspective, which we explain in Appendix. In simulations, the target channel load is set to be $0.6$.

\subsection{Convergence to Target Channel Load}
The evolving equation (\ref{equ:LIMERIC}) shows that in steady state LIMERIC converges to a value strictly smaller than $r_g$ \cite{BanKenRoh_13}. In other words, the target channel load can not be achieved in steady state. However, our algorithm leads to full utilization of the target channel load. See Figure \ref{figure:channelload_compare}. Furthermore, our algorithm converges less than $4$ seconds while in EMBARC oscillations still occur after $10$ seconds.

\subsection{Packet Reception Rate}

We compare the number of successful received packets per second between EMBARC (with $\alpha=0.1$ and $\beta=0.001$) and our joint congestion control algorithm, which motivates the need of congestion control algorithms in DSRC. To be fair with (rate-control only) EMBARC, we simulated our standalone rate control algorithm as well. Figure \ref{figure:NumOf_Received_Packet} shows that:
\begin{enumerate}
\item our rate control algorithm performs uniformly better than EMBARC because of full utilization of the target channel load. Specifically, our rate control guarantees the convergence of measured channel load of each vehicle to the target channel load while EMBARC is proved to have inevitable gap in its steady state (\ref{equ:LIMERIC});
\item the joint congestion control algorithm provides significant gain in short distance regime (safety-sensitive zone).
This is because both rate and transmission range are adjusted according to the deployment topology, as shown in Figure \ref{figure:Joint_performance}. Specifically, the transmission range increases in the sparse segments and achieves maximum at the center vehicle, while the range is constantly short in the dense segments. Note that $~80\%$ vehicles have short range, e.g., $50m$, which leads to the performance gain in the short-range regime.
\end{enumerate}


\subsection{Coverage and Awareness}
\begin{figure}
\begin{centering}
  \includegraphics[width=3.2in]{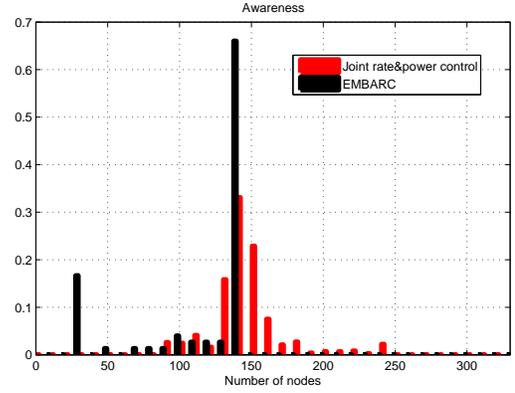}
  \caption{Awareness distribution of joint rate and power control}\label{figure:Joint_Awareness}
  \end{centering}
\end{figure}

\begin{figure}
\begin{centering}
  \includegraphics[width=3.2in]{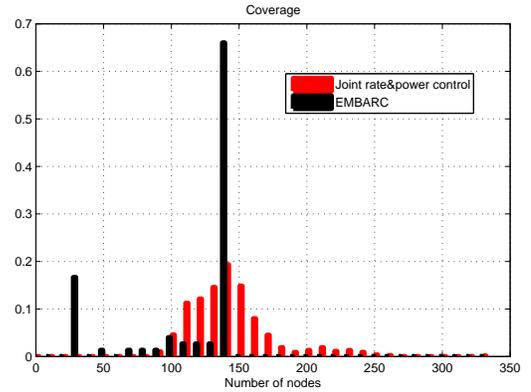}
  \caption{Coverage distribution of joint rate and power control}\label{figure:Joint_Coverage}
  \end{centering}
\end{figure}

%
%
Figures \ref{figure:Joint_Awareness} shows the distribution of the number of vehicles that a vehicle can receive messages from, called {\em awareness}. Figure \ref{figure:Joint_Coverage} shows the distribution of the number of vehicles within a vehicle's transmission range, called {\em coverage}.  Under EMBARC, there are two peaks at $35$ and $145$ in both coverage and awareness, respectively associated with two different densities in the network. The joint algorithm only has one peak since the algorithm dynamically allocates the resources based on the network topology, achieving fairness in terms of both coverage and awareness.

\section{Conclusions}
In this paper, we proposed a utility maximization framework for joint rate and power control in DSRC, and formulated two optimization problems, named Rate-OPT and Power-OPT, where Rate-OPT deals with rate control with fixed transmit power and Power-OPT deals with power control with fixed rates. We developed both distributed rate control and power control algorithms and proved the algorithms are asymptotically optimal. Evaluations using ns2 showed that our algorithms outperform EMBARC at several relevant aspects including channel utilization, packet reception rate, coverage and awareness.

\bibliographystyle{IEEEtran}
\bibliography{v2v}
\newpage

\appendix

\subsection{Proof of Lemma \ref{lem:integer}}
Recall that the optimization problem (\ref{subopt:obj}) is for fixed $i$ and $\lambda.$ Define $$F_i(x,y)=\sum_{j} x_{ij}U_{ij}(\mu_i)-\epsilon \lambda_j\mu_i y_{ij}.$$
Let $(\hat{x}, \hat{y})$ denote an optimal solution. We first have the following two observations:
\begin{itemize}
\item According to constraint (\ref{cons:range2}), $\hat{x}_{ij}=\hat{x}_{ik}$ if $d_{ij}=d_{ik},$ which implies that
\begin{equation}\hat{x}_{ij}=\hat{x}_{ik}\quad \hbox{if}\quad j, k\in {\cal G}_g.\label{eq: range}\end{equation}
\item To maximize the objective (\ref{subopt:obj}), $y$ should be chosen as small as possible. So
\begin{equation}
\hat{y}_{ik}=\max_{j: \beta_{ik}\leq \alpha_{ij}} \hat{x}_{ij}.
\label{eq: sense}
\end{equation} Since $\hat{x}_{ij}\geq\hat{x}_{ik}$ when $\alpha_{ij}<\alpha_{ik},$ (\ref{eq: sense}) is equivalent to
\begin{equation*}
\hat{y}_{ik}=\hat{x}_{ij^\prime} \quad\hbox{where}\quad j^\prime=\arg\min_{j: \beta_{ik}\leq \alpha_{ij}} \alpha_{ij},
\end{equation*} which further implies that
\begin{equation}
\hat{y}_{ik}=\hat{x}_{ij} \quad\hbox{if}\quad k\in{\cal H}_g\hbox{ and } j\in{\cal G}_g.
\label{eq: senseg}
\end{equation}
In other words, $\hat{x}_{ij}$ and $\hat{y}_{ik}$ are equal if $j\in {\cal G}_g$ and $k\in{\cal H}_g$ for the same $g.$ This is easy to understand because the define of H-group ${\cal H}_g$ is the set of vehicles that can sense the transmission of vehicle $i$ when the vehicles in G-group ${\cal G}_g$ can receive messages from vehicle $i.$
\end{itemize}

Now suppose that $\hat{x}_{ij}$ is not an integer solution. Initially, let $\tilde{\hat{x}}=\hat{x}.$ Identify vehicle $j^\prime$ is the vehicle has the maximum $\alpha_{ij}$ among all vehicles such that $0<\hat{x}_{ij}<1,$ i.e.,
$$j^\prime=\arg\max_{j: 0<\hat{x}_{ij}<1} \alpha_{ij}.$$  According to observations (\ref{eq: range}) and (\ref{eq: senseg}), we have $\hat{x}_{ij}=\hat{y}_{ik}=\hat{x}_{ij^\prime}$ for $j\in {\cal G}_g(j^\prime)$ and $k\in{\cal H}_{g(j\prime)}.$ Therefore,
\begin{eqnarray*}
&&\sum_{j\in{\cal G}_{g(j^\prime)}}\hat{x}_{ij}U_{ij}(\mu_i)-\epsilon\sum_{k\in{\cal H}_{g(j^\prime)}}\lambda_k\mu_i\hat{y}_{ik}\\
&=&\left(\sum_{j\in{\cal G}_{g(j^\prime)}}U_{ij}(\mu_i)-\epsilon\sum_{k\in{\cal H}_{g(j^\prime)}}\lambda_k\mu_i\right) \hat{x}_{ij^\prime}.
\end{eqnarray*}
If \begin{eqnarray*}
\sum_{j\in{\cal G}_{g(j^\prime)}}U_{ij}(\mu_i)-\sum_{k\in{\cal H}_{g(j^\prime)}}\lambda_k\mu_i\leq 0,
\end{eqnarray*}
then we define $\tilde{\hat{x}}_{ij}=\tilde{\hat{y}}_{ik}=0$ for $j\in {\cal G}_g(j^\prime)$ and $k\in{\cal H}_{g(j\prime)}.$  Otherwise, we define $\tilde{\hat{x}}_{ij}=\tilde{\hat{y}}_{ik}=\hat{x}_{ib}$ for $j\in {\cal G}_g(j^\prime)$ and $k\in{\cal H}_{g(j\prime)}$ where $b\in {\cal G}_{g(j^\prime)-1}.$ It is easy to see that the following inequality holds: $$F_i(\hat{x}, \hat{y})\leq F_i(\tilde{\hat{x}}, \tilde{\hat{y}}).$$
Now for the second scenario discussed above, we have
\begin{eqnarray*}
&&\sum_{j\in{\cal G}_{g(j^\prime)}\cup{\cal G}_{g(j^\prime)-1}}U_{ij}(\mu_i)\tilde{\hat{x}}_{ij}-\epsilon\sum_{k\in{\cal H}_{g(j^\prime)}\cup {\cal H}_{g(j^\prime)-1}}\lambda_k\mu_i\tilde{\hat{y}}_{ik}\\
&=&\left(\sum_{j\in{\cal G}_{g(j^\prime)}\cup{\cal G}_{g(j^\prime)-1}}U_{ij}(\mu_i)-\epsilon \sum_{k\in{\cal H}_{g(j^\prime)}\cup {\cal H}_{g(j^\prime)-1}}\lambda_k\mu_i\right) \hat{x}_{ib}.
\end{eqnarray*} Similarly, we define $\tilde{\hat{x}}_{ij}=\tilde{\hat{y}}_{ik}=0$ for $j\in {\cal G}_{g(j^\prime)}\cup{\cal G}_{g(j^\prime)-1}$ and $k\in{\cal H}_{g(j\prime)}\cup {\cal H}_{g(j\prime)-1}$ if \begin{eqnarray*}
\sum_{j\in{\cal G}_{g(j^\prime)}\cup{\cal G}_{g(j^\prime)-1}}U_{ij}(\mu_i)-\epsilon\sum_{k\in{\cal H}_{g(j^\prime)}\cup {\cal H}_{g(j^\prime)-1}}\lambda_k\mu_i \leq 0;
\end{eqnarray*} and otherwise define  $\tilde{\hat{x}}_{ij}=\tilde{\hat{y}}_{ik}=\hat{x}_{ic}$ for $j\in {\cal G}_{g(j^\prime)}\cup{\cal G}_{g(j^\prime)-1}$ and $k\in{\cal H}_{g(j\prime)}\cup {\cal H}_{g(j\prime)-1},$ where $c\in {\cal G}_{g(j\prime)-2}.$ Similarly, we also have $$F_i(\hat{x}, \hat{y})\leq F_i(\tilde{\hat{x}}, \tilde{\hat{y}}).$$

Repeating the same argument, we can conclude that there exists $g^\prime$ such that $\tilde{\hat{x}}_{ij}=0$ if $g(j)>g^\prime$ and $\tilde{\hat{x}}_{ij}=x$ if $g(j)\leq g^\prime.$ Therefore,
\begin{eqnarray*}
F_i(\tilde{\hat{x}},\tilde{\hat{y}})=\left(\sum_{j\in\cup_{g:g\leq g^\prime}{\cal G}_{g}}U_{ij}(\mu_i)-\epsilon \sum_{j\in\cup_{g:g\leq g^\prime}{\cal H}_{g}}\lambda_j\mu_i\right)x.
\end{eqnarray*} It is easy to see that we should choose $x=1$ if
\begin{eqnarray*}
\sum_{j\in\cup_{g:g\leq g^\prime}{\cal G}_{g}}U_{ij}(\mu_i)-\epsilon \sum_{j\in\cup_{g:g\leq g^\prime}{\cal H}_{g}}\lambda_j\mu_i>0.
\end{eqnarray*} and $x=0$ otherwise. Therefore, from any optimizer $(\hat{x}, \hat{y})$ we can construct an integer solution $(\tilde{\hat{x}}, \tilde{\hat{y}})$ such that
$$F_i(\hat{x}, \hat{y})\leq F_i(\tilde{\hat{x}}, \tilde{\hat{y}}).$$

From the discussion above, the integer optimizer is
$$x_{ij}=\left\{
           \begin{array}{ll}
             1, & \hbox{if } g(j)\leq g^\prime  \\
             0, & \hbox{otherwise.}
           \end{array}
         \right. \quad\hbox{and}\quad y_{ij}=\left\{
           \begin{array}{ll}
             1, & \hbox{if } h(j)\leq g^\prime  \\
             0, & \hbox{otherwise.}
           \end{array}
         \right.,
$$ where
\begin{align*}
g^\prime=\min&\left\{g: \sum_{j\in\cup_{q:k\leq q\leq g}{\cal G}_{g}}U_{ij}(\mu_i)\right.\\
&\left.-\epsilon\sum_{j\in\cup_{q:k\leq q\leq g}{\cal H}_{g}}\lambda_j\mu_i>0\quad \forall k< g\right\}.
\end{align*}

\subsection{Proof of Theorem \ref{thm:range-opt}}
Defining $V[t]=\sum_j \lambda^2_j[t],$ we have
\begin{eqnarray*}
&&\Delta V[t]\\
&=&V[t+1]-V[t]\\
&\leq&\sum_j \left(\lambda_j[t]+\sum_i {\mathbb I}_{(p_i[t]\geq \beta_{ij}}\mu_i-\gamma\right)^2-\sum_j \lambda^2_j[t]\\
&=&\sum_j \left(\lambda_j[t]+ \sum_i y_{ij}[t]\mu_i-\gamma\right)^2-\sum_j \lambda^2_j[t]\\
&=&\sum_j \left(2\lambda_j[t]+ \sum_i y_{ij}[t]\mu_i-\gamma\right)\left(\sum_i y_{ij}[t]\mu_i-\gamma\right)\\
&=&\sum_j 2\lambda_j[t]\left(\sum_i y_{ij}[t]\mu_i-\gamma\right)+\left(\sum_i y_{ij}[t]\mu_i-\gamma\right)^2.
\end{eqnarray*}
Note that $\left(\sum_i y_{ij}[t]\mu_i-\gamma\right)^2\leq (n\mu_{\max}+ \gamma_{max})^2,$ where $n$ is the maximum number of vehicles whose transmissions can be sensed by vehicle $j,$ since $0\leq y_{ij}[t] \leq 1$ and $0\leq \gamma\leq \gamma_{max}.$

Defining $M^\prime=\sum_i (\gamma_{max}+n\mu_{\max})^2$ and denoting by $(\tilde{x}^*, \tilde{y}^*)$ the optimal solution to problem (\ref{eq:obj1}), we have
\begin{eqnarray*}
\Delta V[t]&=&V[t+1]-V[t]\\
&\leq& M^\prime+2\sum_j \lambda_j[t]\left(\sum_i y_{ij}[t]\mu_i-\gamma\right)\\
&=& M^\prime+2\sum_j \lambda_j[t]\left(\sum_i y_{ij}[t]\mu_i-\gamma\right)\\
&&+\frac{2}{\epsilon}\sum_{i,j} U_{ij}(\mu_i) \tilde{x}^*_{ij}-\frac{2}{\epsilon}\sum_{i,j} U_{ij}(\mu_i) \tilde{x}^*_{ij}\\
&&+\frac{2}{\epsilon}\sum_{i,j} U_{ij}(\mu_i) x_{ij}[t]-\frac{2}{\epsilon}\sum_{i,j} U_{ij}(\mu_i) x_{ij}[t]\\
&&+2\sum_j \lambda_j[t]\left(\sum_i \tilde{y}^*_{ij}\mu_i-\gamma\right)\\
&&-2\sum_j \lambda_j[t]\left(\sum_i \tilde{y}^*_{ij}\mu_i-\gamma\right).
\end{eqnarray*} By rearranging the terms above, we obtain
\begin{align}
&\Delta V[t]\leq M^\prime\\
&-\frac{2}{\epsilon}\sum_{i,j} U_{ij}(\mu_i) x_{ij}[t]+2\sum_j \lambda_i[t]\left(\sum_i y_{ij}[t]\mu_i-\gamma\right)\label{eq:l3}\\
&+\frac{2}{\epsilon}\sum_{i,j} U_{ij}(\mu_i) \tilde{x}^*_{ij}-2\sum_j \lambda_i[t]\left(\sum_i \tilde{y}^*_{ij}\mu_i-\gamma\right)\label{eq:l4}\\
&-\frac{2}{\epsilon}\sum_{i,j} U_{ij}(\mu_i) \tilde{x}^*_{ij}+\frac{2}{\epsilon}\sum_{i,j} U_{ij}(\mu_i) x_{ij}[t]\\
&+2\sum_j \lambda_i[t]\left(\sum_i \tilde{y}^*_{ij}\mu_i-\gamma\right). \label{eq:l6}
\end{align}

According to Lemma \ref{lem:integer}, $(\ref{eq:l3})+(\ref{eq:l4})\leq 0.$ Further, since $(\tilde{x}^*, \tilde{y}^*)$ is a feasible solution to problem (\ref{eq:obj1}), we have $(\ref{eq:l6})\leq 0.$ Hence, we conclude
\begin{eqnarray*}
\Delta V[t]\leq M^\prime-\frac{2}{\epsilon}\sum_{i,j} U_{ij}(\mu_i)\tilde{x}^*_{ij}+\frac{2}{\epsilon}\sum_{i,j} U_{ij}(\mu_i) x_{ij}[t],
\end{eqnarray*} which implies
\begin{eqnarray*}
\frac{\epsilon}{2} (\Delta V[t]-M^\prime)+\sum_{i,j} U_{ij}(\mu_i) \tilde{x}^*_{ij}\leq \sum_{i,j}U_{ij}(\mu_i) x_{ij}[t],
\end{eqnarray*}
Note that $\sum_{t=0}^{T-1}\Delta V[t]=V[T]-V[0],$ so
\begin{eqnarray*}
&\frac{\epsilon}{2T} \left(V[T]-V[0]\right)-\frac{\epsilon M^\prime}{2}+\sum_{i,j} U_{ij}(\mu_i) \tilde{x}^*_{ij}\\
\leq& \frac{1}{T}\sum_{t=0}^{T-1}\sum_{i,j} U_{ij}(\mu_i) x_{ij}[t],
\end{eqnarray*} which implies
\begin{eqnarray*}
-\frac{\epsilon V[0]}{2T} -\frac{\epsilon M^\prime}{2}+\sum_{i,j}U_{ij}(\mu_i)\tilde{x}^*_{ij}\leq \frac{1}{T}\sum_{t=0}^{T-1}\sum_{i,j} \frac{\eta_{ij}}{d_{ij}} x_{ij}[t].
\end{eqnarray*} The theorem holds by choosing $M=M^\prime/2$ and because $$\sum_{i,j}U_{ij}(\mu_i) \tilde{x}^*_{ij}\geq \sum_{i,j} U_{ij}(\mu_i) {x}^*_{ij}.$$

\subsection{Selection of Utility Function (\ref{eq: utility.example})}
Note that given distance $d_{ij}$ and relative speed $v_{ij},$ vehicles $i$ and $j$ on a line would collide after $\frac{d_{ij}}{v_{ij}}$ units of time if they do not change their speeds. We therefore call $\frac{d_{ij}}{v_{ij}}$ {\em reaction time of pair $(i,j).$}  To prevent the collision, vehicles $i$ and $j$ need to communicate at least once during this reaction time. Assume each message is reliably received with probability $p,$ then the probability that vehicle $j$ receives at least one message from vehicle $i$ during the reaction time is $$1-(1-p)^{\frac{d_{ij}}{v_{ij}}\mu_i}.$$ Imposing a lower bound $p_{\min}$ on this probability is equivalent to requiring that
\begin{equation}
\frac{d_{ij}}{v_{ij}}\mu_i\geq \frac{\log(1-p_{\min})}{\log(1-p)}.\label{eq: safety}
\end{equation}
These pairwise safety constraints may not always be feasible depending on the network density and the geographical distribution of the vehicles.

\begin{figure}[htb]
\begin{centering}
  \includegraphics[width=3in]{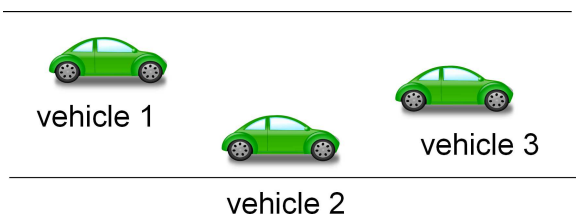}
  \caption{Vehicle 1 and vehicle 2 are both approaching vehicle 3 with different speeds}\label{figure:example1}
  \end{centering}
\end{figure}

Therefore, we consider a different requirement also from the collision avoidance perspective. Consider a scenario shown in Figure \ref{figure:example1}, where both vehicle $1$ and vehicle $2$ are approaching vehicle $3$ with different speeds. A fair resource allocation should equalize the reaction time to avoid collisions, i.e., to have \begin{equation}\frac{d_{13}}{v_{13}}\mu_1=\frac{d_{23}}{v_{23}}\mu_2.\label{eq:equal}\end{equation} In a general scenario, this objective could also be difficult to achieve. We note that
if we assume all vehicles share a single-bottleneck link, e.g., in the scenario in Figure \ref{figure:example1} where all vehicles can hear each other, then solving the following optimization problem
\begin{equation}\max_{\mu} \frac{v_{13}}{d_{13}}\log \mu_1+\frac{v_{23}}{d_{23}}\log \mu_2\end{equation} results in the solution
\begin{equation}\frac{v_{13}}{d_{13}}\frac{1}{\mu_1}=\frac{v_{23}}{d_{23}}\frac{1}{\mu_2},\end{equation} which is equivalent to (\ref{eq:equal}). This motivated the objective function in (\ref{eq: utility.example}), where each link $d_{ij}$ is associated with a weighted log-utility $\frac{\max\{v_{ij},\alpha\}}{d_{ij}}\log {\mu_i}.$ Since $v_{ij}$ may be negative, we define the weight to be $\max\{v_{ij}, \alpha\}$ for some $\alpha>0.$

\end{document}